\documentclass[showpacs,twocolumn,prl]{revtex4}%
\usepackage{amsmath}
\usepackage{graphicx}
\usepackage{bm}
\usepackage{amsfonts}
\usepackage{amssymb}%
\begin{document}
\title{Intrinsic Spin Hall Edges}
\author{\.{I}nan\c{c} Adagideli}
\altaffiliation{current address: \textit{Department of Physics and Astronomy,
University of British Columbia, 6224 Agricultural Road, Vancouver, B.C. V6T 1Z1, Canada} }
\affiliation{Kavli Institute of Nanoscience, TU Delft, Lorentzweg 1, 2628 CJ Delft, The Netherlands}
\author{Gerrit E.W. Bauer}
\affiliation{Kavli Institute of Nanoscience, TU Delft, Lorentzweg 1, 2628 CJ Delft, The Netherlands}
\date{\today}

\begin{abstract}
The prediction of intrinsic spin Hall currents by Murakami \textit{et al.}
and\ Sinova \textit{et al.} raised many questions about methods of detection
and the effect of disorder. We focus on a contact between a Rashba type spin orbit
coupled region with a normal two-dimensional electron gas and show that
the spin Hall currents, though vanishing in the bulk of the sample, can be recovered
from the edges. We also show that the current induced spin accumulation in the spin
orbit coupled system
diffuses into the normal region and contributes to the spin current in the leads.
\end{abstract}

\maketitle

Transport and manipulation of spins in semiconductor structures
has become a mainstream in condensed matter physics~\cite{Zutic}.
In principle, spins can be injected into semiconductors by
ferromagnets via electric contacts. However, finding suitable
material combinations that do not suffer from the conductance
mismatch~\cite{Schmidt}, turned out to be difficult. Furthermore,
introducing ferromagnetic materials into the semiconductor
microfabrication process is undesirable from a technological point
of view. The prospect to generate spin accumulations in
semiconductors without ferromagnets or applied magnetic fields
simply by driving a current through a material with intrinsic
spin-orbit (SO) interaction and broken
inversion-symmetry~\cite{Levitov,Edelstein,Katoaccum} is therefore
very attractive. A related effect that attracted a lot of
attention is the spin Hall effect (SHE), \textit{i.e}.~the spin
current(SC) that has been predicted to flow normal to an applied
electric current in the absence of an applied magnetic field. When
caused by impurities with spin-orbit
scattering~\cite{DP,extrinsic} this effect is called
\textquotedblleft extrinsic\textquotedblright. A spin Hall current (SHC)
can also be generated by the spin-orbit interaction of the lattice
potential as has recently been predicted for p-doped III-V
semiconductors~\cite{Murakami} and the two-dimensional electron
gas with a Rashba-type SO interaction (R2DEG)~\cite{Sinova}.
Whether the experimental observations of the spin Hall effect by
optical methods~\cite{exp} have intrinsic or extrinsic origin is
still a matter of debate. In spite of initial controversies, analytic
theories~\cite{Inoue,pdoped,Mishchenko,Khaetskii,Chalaev} as well as
numerical simulations \cite{Nomura,Sheng} consistently predict that the
SHE should vanish in the disordered (bulk) R2DEG~\cite{SHE_Workshop}.
Some doubts remain whether the SC, being a non-conserved quantity in SO
coupled systems, is observable at all~\cite{Rashba}.

In this Letter, we focus on the spin currents near normal contacts. First, an elementary
and general proof is given that the spin Hall effect due to the lattice SO
coupling (\textit{viz.}~intrinsic SHE) must vanish in diffuse bulk systems
with an arbitrarily strong SO interaction that is linear in the electron wave
vector. Nevertheless, using an extension of this argument to finite system sizes,
we show that near the edges a spin Hall current can persist.
Next, by solving the kinetic equations for a model system of a R2DEG in contact
with a normal metal system without SO interaction, we calculate indeed a finite
SHC.
This SC is generated in a skin
depth determined by the Dyakonov-Perel~\cite{DP} spin-flip diffusion length ($L_s$)
and the polarization is not normal to the 2DEG, having a component due to the
diffusion current from the SO-generated spin accumulation (SA).
The magnitude of the SC generated at the edges
depends on whether the system is clean (impurity broadening less than the SO
splitting) or dirty (opposite limit). However, in contrast to the bulk SC, the edge SC
does not vanish when the system is not ballistic ($L_s$ smaller than
the system size).
The SC is calculated in the normal metal contact and therefore certainly a
transport current~\cite{Rashba}.
Related work on interface and boundary effects focused so far on mesoscopic systems via numerical
simulations~\cite{Nikolic} and the SA near hard wall boundaries~\cite{hardwall}.

We proceed to derive a transport equation valid in the Boltzmann limit that is
capable of handling the full spin dynamics.
In $2\times2$ spin space, the Hamiltonian is
\begin{equation}
H=\mathbf{p}^{2}/{2m}+V(\mathbf{x})+H^{R}-e{\bm E}(t)\cdot\mathbf{x}%
,\label{EQ:hamiltonian}%
\end{equation}
where $\mathbf{x}$ and $\mathbf{p}$ are the (two-dimensional) position and
momentum operators, respectively. Here the unit vector $\hat{\bm z}$ is normal to
the 2DEG, $H^{R}=(\alpha/\hbar)\mathbf{p}\cdot({\bm\sigma}\times\hat{\bm z})$
is the Rashba Hamiltonian with Pauli matrices ${\bm\sigma}$ and $\alpha$
parameterizes the strength of the SO interaction~\cite{Nitta}, ${\bm E}$ is
the electric field, and $V(\mathbf{x})=\sum_{i=1}^{N}\phi(\mathbf{x}-{\bm
X}_{i})$ is the impurity potential, modelled by $N$ impurity centers located
at points $\{{\bm X}_{i}\}$. Although it is possible to consider ac fields, we
focus here on dc fields in the $x-$direction and assume that the electric
field is turned on adiabatically in the remote past at which the system was in
thermal equilibrium, \textit{i.e}.~we assume ${\bm E}=\lim_{s\rightarrow0}{\bm
E}_{0}\exp(st/\hbar)$. To leading order in the impurity potential the diagonal
elements of the density matrix in reciprocal space satisfy the following
equation~\cite{Luttinger}:
\begin{align}
-isf({\bm k})+[f({\bm k}),H_{k}^{R}] &  =\sum_{k^{\prime}}\Big(f_{\bm k\bm
k^{\prime}}V_{\bm k^{\prime}\bm k}-V_{\bm k\bm k^{\prime}}f_{\bm k^{\prime}\bm
k}\Big)\nonumber\\
&  \quad+\,\,e{\bm E}\cdot\lbrack{\bm x},f^{0}].\label{EQ:Diag}%
\end{align}
Here $f^{0}=(f_{+}^{0}+f_{-}^{0})/2+\sigma_{\theta}(f_{+}^{0}-f_{-}^{0})/2$
is the equilibrium density matrix with $f_{\pm}^{0}(k)=\mathcal{F}%
\big(\hbar^{2}k^{2}/2m\pm\alpha k\big)$, where $\mathcal{F}(E)$ is the Fermi
function and $\sigma_{\theta}=\mathbf{k}\cdot({\bm\sigma}\times{\bm z})/k$.
The off-diagonal elements of the density matrix read
\begin{equation}
f_{\bm k\bm k^{\prime}}=\frac{i}{2\pi}\int_{-\infty}^{\infty}dE\,G_{\bm k}%
^{R}(E)\big(f({\bm k})-f({\bm k}^{\prime})\big)G_{{\bm k}^{\prime}}%
^{A}(E)\,V_{\bm k\bm k^{\prime}}.\label{EQ:ODiag}%
\end{equation}
Here $G_{k}^{R(A)}(E)=(E-H_{0}-H_{k}^{R}+(-)is/2)^{-1}$ are retarded
(advanced) matrix Green functions. Substituting Eq.~(\ref{EQ:ODiag}) into
Eq.~(\ref{EQ:Diag}) and averaging over $V_{kk^{\prime}}$ (in the Boltzmann
limit averaging is equivalent to replacing $|V_{kk^{\prime}}|^{2}$ with its
average value $N|\phi_{kk^{\prime}}|^{2}/A^{2}$, where $N$ is the number of
impurities and $A$ is the area~\cite{Luttinger}) gives our basic equation,
valid for weak $V$ and low enough impurity densities to ignore weak
localization effects, but to all orders in $\alpha$.

The mechanism behind the intrinsic SHE is the spin precession of
quasiparticles while being accelerated by the electric field~\cite{Sinova}.
However, impurity scattering provides a brake that in the steady state cancels
the acceleration on average. Therefore the SHE should vanish in an infinite,
homogeneously disordered system. This idea can be formally expressed by
considering the acceleration operator
$\mathbf{\ddot{x}}_{i}=e{\bm E}_{i}/m-\nabla_i V/m
-2\alpha^{2}\hbar^{-3}\epsilon_{3ji}\sigma_{3}\mathbf{p}_{j}$,
where $\epsilon_{ijk}$ is the antisymmetric tensor and the Einstein summation
convention is implied. We notice that the last term is proportional to the
$j$'th component of the SC operator polarized in the ${z-}%
$direction, $J_{j}^{z}=\left\{  {\bm v}_{j},{\bm\sigma}_{z}\right\}  $. The
expectation value is defined by $\langle O\rangle\equiv\overline
{\mathrm{Tr}fO}$, where the trace is over wave vector and spin space, and
$\overline{\cdots}$ denotes averaging with respect to impurity configurations.
In a steady state the average acceleration $\left\langle \mathbf{\ddot{x}}%
_{i}\right\rangle $ must vanish, leading to the equality
\begin{equation}
2\alpha^{2}m^{2}\hbar^{-3}\,\epsilon_{3ji}\langle J_{j}^{z}%
\rangle=e{\bm E}_{i}-\langle\nabla_i V\rangle
.\label{EQ:spcrnt}%
\end{equation}
We show that the right hand side of this equality also vanishes by evaluating
the expectation value of the deceleration due to impurity scattering:
\begin{align}
\langle\nabla_i V\rangle &  =-i\sum_{kk^{\prime}}({\bm
k}_{i}-{\bm k}_{i}^{\prime})V_{kk^{\prime}}\mathrm{tr}f_{{\bm k}^{\prime}{\bm
k}}\nonumber\\
&  =-i\sum_{k}\mathrm{tr}\left({\bm k}_{i}[f({\bm k}),H_{{\bm k}}^{R}]+i{\bm
k}_{i}e{\bm E}_{j}
\nabla_{k_j} f^{0}
\right)
=e{\bm E}_{i}, \nonumber
\end{align}
where $\mathrm{tr}$ is the trace over spin components and Eqs.~(\ref{EQ:Diag}%
-\ref{EQ:ODiag}) have been used in the second step. Substituting the expression
above into Eq.~(\ref{EQ:spcrnt}) we see that all components of the SC polarized
in the $z-$direction vanish with the average acceleration~\cite{caveat}.
This result holds for infinite systems regardless of the range of the impurity potential
or whether the system is clean ($\alpha k_F \tau/\hbar\gg 1$) or dirty
($\alpha k_F \tau/\hbar\ll 1$), where $\tau$ is the momentum lifetime. Thus
generalizing previous results~\cite{Inoue,pdoped,Mishchenko,Khaetskii,Chalaev}.
However, as we shall show below, for semi-infinite and finite systems, SCs persist
near the edges, but the size of these currents depend on whether the
system is clean or dirty.

This line of argument allows one to check related Hamiltonians. In the presence of $k$-linear
Dresselhaus and Rashba terms, the result remains unchanged besides the
substitution $\alpha^{2}\rightarrow\alpha^{2}-\beta^{2}$, where $\beta$ is the
Dresselhaus spin orbit coupling constant. Thus the SHC still
vanishes (with the possible exception of the degeneracy point $\alpha=\beta$
\cite{FT:dresselhaus}). When the SO coupling contains cubic terms
like $\alpha(k)=\alpha_{0}+\alpha_{1}k^{2}$, it is easy to show that the SHC
is proportional to $\alpha_{1}$~\cite{pdoped}. Another possible
situation is the presence of a Zeeman field: in this case the operator
equation is modified to give $\alpha\epsilon_{3ji}\langle J_{j}^{z}%
\rangle=\langle\sigma_{i}\rangle B_{3}-\langle\sigma_{3}\rangle B_{i}$,
relating the SHC to the SA. If $\alpha$ varies
in space, the SHC is found to be proportional to the spatial
derivatives of $\alpha$ and $f$. If $\alpha$ is constant, but $f$ varies,
\textit{e.g}. due to boundaries or interfaces~\cite{Mishchenko}%
, the SHC is proportional to the gradients of the density
matrix:
\begin{equation}
\frac{4\alpha^{2}m^{2}}{\hbar^{3}}\,\epsilon_{3ji}\langle\hat{J}_{j}%
^{z}\rangle=\langle k_{i}\{\frac{\hbar^{2}k_{l}}{m}+\alpha(\hat{\bm z}%
\times\boldsymbol{\sigma})_{l},\nabla_{l}f({\bm k},{\bm x})\}\rangle
.\label{EQ:scid}%
\end{equation}
This equation shows that although the bulk SH current vanishes, there is no a priori
reason for SH currents near the edges of the R2DEG to vanish. Next, we shall
show indeed the SH currents do not vanish near the edges. We
therefore return to the quantum transport equation, allowing for spatially
varying density matrices but assuming short-range s-wave scatterers with
$\overline{|V_{kk^{\prime}}|^{2}}=N\lambda^{2}/A$ and $\alpha/k_{F}\ll1$. We
solve the transport equation by expressing $f$ in terms of a gradient
expansion of $\rho(E)=(i\hbar^{2}/2\pi m)\sum_{k}\big(G_{k}^{R}(E)f({\bm
k})-f({\bm k})G_{k}^{A}(E)\big)$. In the case of s-wave scatterers
and to leading order in $m\alpha/\hbar^2 k_{F}$, this generates the same diffusion
equation as Ref.~\cite{Mishchenko} and Burkov \textit{et al}. in
Ref.~\cite{Sinova}. In terms of components of the density matrix:
$\rho=n+\mathbf{s}\boldsymbol{\cdot\sigma}+\sigma_3 s_3$ :
\begin{align}
D\nabla^{2}n-4K_{s-c}(\boldsymbol{\nabla}\times\mathbf{s})_{z} &
=0\label{diff}\\
D\nabla^{2}s_{3}-2K_{p}(\boldsymbol{\nabla}\cdot\mathbf{s}) &  =\frac{2s_{3}%
}{\tau_{s}}\\
D\nabla^{2}\mathbf{s}+2K_{p}\boldsymbol{\nabla} s_{3}-K_{s-c}(\mathbf{z}\times\boldsymbol{\nabla})n
&=\frac{\mathbf{s}}{\tau_{s}}
\end{align}
Here $D=v_{F}^{2}\tau/2$, $\tau_{s}=\tau(1+4\xi^{2})/2\xi^{2}$, $K_{s-c}%
=\alpha\xi^{2}/(1+4\xi^{2})$, $K_{p}=\hbar k_{F}\xi/m(1+4\xi^{2})^{2}$ and
$\xi=\alpha k_{F}\tau/\hbar$.
The SC is given, in the diffuse limit by
\begin{equation}
j_j^i=\frac{v_F \xi }{1+4\xi^2}\left(
\delta_{i3}\left(s_j-\epsilon_{jm3}\frac{\alpha\tau}{2}\nabla_m n\right)
-\delta_{ij}s_3\right)-D\nabla_j s_i.
\label{EQ:SCexp}
\end{equation}
Electric field dependence can be reintroduced by the
substitution $\boldsymbol{\nabla}\rightarrow\boldsymbol{\nabla}+e{\bf E}\partial_E$.

\begin{figure}[ptb]
\mbox{ \includegraphics[width=8cm]{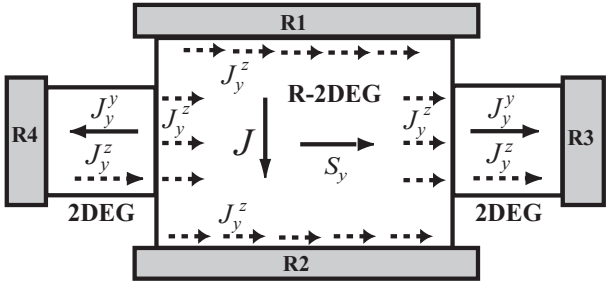}}
\caption{The schematic setup for SC generation in an R2DEG that is current
biased by reservoirs R1 and R2 and equipped with 2DEG Hall
contacts with vanishing spin orbit coupling to reservoirs R3 and
R4. The latter can be magnetic or nonmagnetic and voltage biased
such that no charge current flows through the 2DEGs. The dashed
arrows indicate the local SHC densities that are
concentrated near the interfaces. The dominant part of the spin
\textit{current\/} flowing into R3 (or R4) is generated within a
skin depth $L_s$ near the corresponding
interface and contains a diffusion term from the SA
$S_{y}$ in the bulk of the R2DEG. The contribution of the SHC density near R1 and R2 to the net
SC~\cite{Mishchenko} flowing into R3 (or R4) is exponentially small.}%
\label{FIG:system}%
\end{figure}

We now focus on a four terminal structure as depicted in Fig.~\ref{FIG:system}.
This structure consists of two massive reservoirs biased to produce a charge
current in the $x-$direction. Between the reservoirs there is a R2DEG hybrid
structure, with $\alpha(\mathbf{x})=\alpha_{0}$ for $L>y>0$ and $\alpha
(\mathbf{x})=0$ for $y<0$ and $y>L$, and additional differences are
disregarded.
The normal 2DEGs are coupled to massive reservoirs R3 and R4,
that are biased such that the charge current is zero, but a SC can still be
collected. To the order (in $\alpha/k_F$) that we are considering,
$n$ does not depend on $y$. Alternatively, one can assume that the transverse
size of the leads to R3 and R4 are much smaller
than the distance between R1 and R2, in which case one can also neglect the $y$ dependence
of $n$.
We consider the distributions at a safe distance from the
reservoirs R1 and R2. In this region $s_{y}$ and $s_{z}$ depend only on
$y$ and $s_{x}=0$ and the diffusion equation becomes:
\begin{equation}%
\begin{pmatrix}
\frac{d^{2}}{d\bar{y}^{2}}-1 & 2\eta\frac{d}{d\bar{y}}\\
-2\eta\frac{d}{d\bar{y}} & \frac{d^{2}}{d\bar{y}^{2}}-2
\end{pmatrix}%
\begin{pmatrix}
s_{2}\\
s_{3}%
\end{pmatrix}
=%
\begin{pmatrix}
\tau_{s}K_{s-c}\frac{dn}{dx}\\
0
\end{pmatrix}
,
\end{equation}
where $\eta=
(1+4\xi^{2})^{-3/2}$, $\bar{y}=y/L_{s}$
and $L_{s}=\sqrt{D\tau_{s}}$.

In order to derive the matching condition for the spin and charge distribution
functions at the contacts (\textit{i.e.} interface between the R2DEG and 2DEG),
short-range fluctuations of boundaries and interfaces that can lead to additional
spin relaxation~\cite{REF:Vasko} are disregarded. We consider an
arbitrary solution $\chi$ of the Schr\"{o}dinger equation set by the
Hamiltonian Eq.~(\ref{EQ:hamiltonian}). We label the solutions in the R2DEG
and 2DEG regions $\chi_{R}$ and $\chi_{N}$, respectively.
At the interface
$\chi_{R}|_{0}=\chi_{N}|_{0}$ and $\mathbf{n}\cdot(i\nabla+\alpha
_{0}(\mathbf{z}\times\sigma))\chi_{R}|_{0}=i\mathbf{n}\cdot\nabla\chi_{N}%
|_{0},$ where $\mathbf{n}$ is the unit vector normal to the interface.
Multiplying from the left with $\chi|_{0}^{\dag}\,\sigma_{i}$ and evaluating
the imaginary part we obtain:
\begin{align}
&  \mathbf{n}\cdot\Big(i\chi_{N}^{\dag}\sigma_{i}\nabla\chi_{N}-i(\nabla
\chi_{N})^{\dag}\sigma_{i}\chi_{N}\Big)\Big|_{0}\nonumber\\
&  \qquad\qquad\qquad=\mathbf{n}\cdot\Big(i\chi_{R}^{\dag}\sigma_{i}\nabla
\chi_{R}-i(\nabla\chi_{R})^{\dag}\sigma_{i}\chi_{R}\nonumber\\
&  \qquad\qquad\qquad\qquad\qquad+\alpha_{0}\chi_{R}^{\dag}\{\sigma
_{i},(\mathbf{z}\times\sigma)\}\chi_{R}\Big)\Big|_{0}\;.\nonumber
\end{align}
We identify the right (left) hand side of this equation as the SC density in
the Rashba (normal) 2DEG. In terms of spin density matrices we have
${\rm tr}\big(f^R\{\sigma_i,{\bf n}\cdot {\bf j}(0)\}\big)=%
{\rm tr}\big(f^N\{\sigma_i,{\bf n}\cdot {\bf j}(0)\}\big)$, where
${\bf j}({\bf x})\equiv \{\mathbf{v},\delta(\hat{\bf x}-{\bf x})\}$ is the local current
density operator.
We therefore have to match the normal components of
the SC density given by Eq.~(\ref{EQ:SCexp}) at the interface~\cite{softBC}.
Since the operator $\{\mathbf{v}%
_{i},\boldsymbol{\sigma}_{j}\}$ can have a nonzero expectation value in the
equilibrium state it has been questioned whether it governs transport of spins
in the presence of SO\ interaction \cite{Rashba}. We notice that the negative
energy solutions (relative to the band crossing) of the Hamiltonian
Eq.~(\ref{EQ:hamiltonian}) without the electric field term, are localized to
the R2DEG region if surrounded by a region with $\alpha=0$. In a normal 2DEG
surrounding the R2DEG, we can therefore show that equilibrium SCs exposed in
Ref.~\cite{Rashba} do not transmit into the normal region. Moreover, the
expectation value of the SC density operator vanishes for these localized
solutions and it is precisely the absence of contributions from these solutions
that shifts the equilibrium value of the SC to zero.

Returning to the setup in Fig.~\ref{FIG:system}, we assume that the reservoirs
R3 and R4 are sufficiently large such that all components of the SA at their
respective interfaces with the ordinary 2DEG leads
vanish. Shrinking the widths of the 2DEGs to zero, we obtain the effective
boundary conditions $s_i=0$ at the R2DEG$|$R3 interface. The finite Ohmic
resistance of a finite 2DEG region between the R2DEG
and the reservoir can easily be reintroduced if necessary and would lead to
somewhat smaller spin conductances. We then can solve the diffusion equation
above and obtain the spin current using Eq.~(\ref{EQ:SCexp}).
\begin{figure}[ptb]
\begin{center}
\includegraphics[width=8cm]{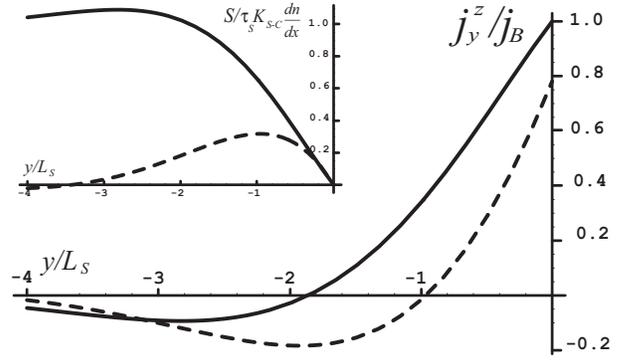}
\end{center}
\caption{SCs and accumulations in the R2DEG ($y<0$) as a function of distance
from the interface to the 2DEG. The full line is the local value of the SHC
density for $\alpha k_{F}\tau/\hbar=0.1$. At the boundary the
SHC density recovers its maximum value
$j_{B}=eEK_{s-c}/2\pi\alpha$
(it approaches the universal value $eE/8\pi$ in the clean limit $\xi\gg1$). The dashed line is the
$z$-component of the diffusion current density. Inset: Corresponding local SA
in the same system normalized to the magnitude of the bulk SA
$eE\alpha\tau m/2\pi$. The solid and dashed lines represent the $y$ and
$z$ components, respectively.}%
\label{FIG:NSOInt}%
\end{figure}

Analytical formulae turn out to be too lengthy to reproduce here. Our
results are therefore summarized in Fig.~\ref{FIG:NSOInt}. The SA is suppressed
at the interface, reflecting the massive-reservoir
boundary condition. The gradient of the two components $s_{y}$ and $s_{z}$
leads to two SC components. The SC polarized in the $y$-direction represents
the out-diffusion of the bulk $s_{y}$ SA. This is not a Hall
current, since it flows into the side contacts with opposite directions
(Fig.~\ref{FIG:NSOInt})
with polarization that is inverted with the bias current direction.
The resulting spin conductivity at the interface is
$\sigma_{xy}^{y}=0.87e\xi^{2}/2\pi$ in the dirty limit($\xi\ll 1$). For larger values of
$\xi$, $\sigma_{xy}^{y}$ increases above this quadratic behavior, but in the
clean limit ($\xi\gg1$) this increase is cut off by the resistance of the normal region.
When spins diffuse from a finite distance into the 2DEG, they precess in the
SO-generated magnetic field. Consequently there is a
diffusion current polarized along the $z$-direction, for which we find in the
dirty limit a conductivity $\sigma_{xy}^{z}=0.83e\xi^{2}/2\pi$. The
conductivity $\sigma_{xy}^{z}$, contrary to $\sigma_{xy}^{y}$,
\textit{decreases} below this quadratic behavior for larger values of $\xi$
and vanishes in the clean limit. Nikolic \textit{et al}.~\cite{Nikolic}
recently observed SCs with $z$ and $y$ polarization in numerical
simulations. In addition, we also find a SHC exponentially
localized to the edges that decays in the bulk on the length scale $L_s$
and reaches its maximum value $eE K_{s-c}/2\pi\alpha$ at the interface to the
reservoir. This is due to the fact that the first term in Eq.~(\ref{EQ:SCexp})
being proportional to $s_y$(thus zero at the interface) is no longer screening
the second term (proportional to $\nabla n$) and reflects the physical process
that the SHC density generated near the interface can escape into
the reservoir before it decays due to spin relaxation.
The resulting dc
spin Hall conductivity is given by $e\xi^{2}/2\pi$ in the dirty and $e/8\pi$
in the clean limit. Our result differs from that of Ref.~\cite{Mishchenko} who
did not take into account the edge currents and obtained similar values only
for the ac response at carefully tuned frequencies.

In conclusion, we find that in a Hall geometry two different spin currents can
be extracted by the Hall contacts from the current-biased disordered R2DEG.
In addition to the SHC, the current-induced SA
drives a spin-diffusion current. The SO generated spin
accumulation is therefore not confined to the region where it is generated,
but can be extracted and, at least in principle, used as a source of spins for
spintronics applications. Both diffusion and SHCs are generated
within a strip that scales like the Dyakonov-Perel spin diffusion length.

We thank Junichiro Inoue, Philip Stamp, Fei Zhou and especially Yuli V.
Nazarov for useful discussions. This work was supported by the FOM, EU
Commission FP6 NMP-3 project 05587-1 \textquotedblleft SFINX\textquotedblright%
, NSERC Canada discovery grant number R8000 and PITP.


\begin{thebibliography}{99}
\bibitem{caveat}
Refs.~\cite{Chalaev} show that the SH current operator is related to the
time derivative of the magnetization (rather than the acceleration).
The SHE then vanishes for stationary SAs. We require only that
the charge current be stationary. For instance, near boundaries the SA has been
observed to precess at constant charge current.~\cite{exp}
\bibitem{softBC}
Our matching conditions also hold for non-abrupt interfaces with varying $\alpha$, provided
the thickness of the interface region is sufficiently smaller than the DP spin relaxation
length.
\end{thebibliography}
\end{document}